\documentclass[prb]{revtex4}
\usepackage{epsf}
\begin{document}
\title{Magic and Mystery  of Mathematics}
\author{Patrick Das Gupta}
\affiliation{Department of Physics and Astrophysics, University of Delhi, Delhi - 110 007 (India)}
\email{patrick@srb.org.in}
\begin{abstract}
Mathematics is probably the only subject that can be classified both as art as well as  science - former, because it is not constrained by the real world and latter because  it is a logical system with precisely defined rules as well as primitives that lead to unambiguous  nontrivial theorems. Indian (and of Indian origin) mathematicians  have continued to do seminal work till present times, culminating in Manjul Bhargava receiving the Fields Medal last year. In such fabulous times, a non-mathematician ponders about the nature of mathematics, and revisits the question: why are fundamental laws of Nature  inherently mathematical?    
\end{abstract}
\maketitle
\section*{`Because it is there' of Mathematics}
With Manjul Bhargava winning the prestigious Fields Medal, Subhash Khot bagging the Rolf Nevanlinna Prize and Ashoke Sen receiving the Dirac Medal,  all in 2014, mathematics has become a source of  non-trivial excitement among the young in the land  of Srinivasa Ramanujan. Ashoke Sen, an internationally renowned string theorist, is not technically a mathematician. But that is a matter of little consequence since string theory, though a branch of physics, is almost inseparable from proper subsets of  advanced pure mathematics. To gauge its importance to  mathematics, one may recollect that   Edward Witten,  {\it numero uno} of string theory and a  stalwart of theoretical physics, was awarded Fields Medal   in 1990 for his seminal  papers  on    supersymmetry, Morse and Hodge-de Rham theories,   that led to  significant progress in number theory and complex analysis  \cite{m1}.

Since antiquity, India has continued to produce many eminent and talented mathematicians like Aryabhatta, Brahmagupta, Bhaskara, Nilakantha,..., Ramanujan, Harish-Chandra, etc. till present times \cite{m5}.  But why is it that a fraction of gifted people, however  minuscule,  keep scaling  newer  heights  of an  esoteric and abstract subject called  mathematics? A  George Mallory type  answer `because it is there' notwithstanding, one of  the reasons could be  the inherent affinity in many for  tackling riddles, whodunnits and paradoxes. 

Some of us may recall  being deeply immersed  time to time    in resolving  conundrums associated with  either unexpected-hanging puzzle or   Klein bottles or Mobius band from the Late Martin Gardner's feature `Mathematical Games’ (which, later,  made a metamorphosis  to `Metamagical Themas’, an anagram of the original title, in the deft hands of Douglas Hofstadter, after he took charge of the column) that appeared  in every issue of the {\it Scientific American} magazine from mid-50s to 80s.  This monthly  feature  initiated several young students  to  logical paradoxes like  Bertrand Russell's alluring  poser   involving a fictitious village consisting  of  barbers who shave only  those men who do not shave themselves. Should a barber, belonging to  this set, shave himself? As young students, we found ourselves  tying  into knots  fathoming  this one! Mathematical theory of knots was of no use in extricating ourselves from this tangled web. 
  
  Russell's paradox had  serious implications in  axiomatic set theory \cite{m2}. It paved the way to sharpen concepts in set theory,  a field which forms one of the cardinal pillars in all branches of mathematics. It will not be a heresy to state that every branch of mathematics begins with sets and mappings.
 
\section*{Roots}

Inception of  nascent topics of mathematics  like arithmetic and geometry  can be attributed to  necessity, factoring in human evolution and survival. 
 Archeological evidence  suggest that  Indus Valley people of $\sim $ 2500 - 1900 BCE used vertical strokes to represent numbers \cite{m3}. Early humans  not only had to keep  track of  their possessions through  counting  but  also needed to estimate directions, distances, shapes and sizes, without which hunting, exploration, building cities like that of Harappa and  Mohenjo daro, raising humongous  Egyptian pyramids, etc, would not have been possible.

 Homo sapiens who could assess numbers and sizes, and discern shapes and directions, had  an evolutionary advantage  for survival during the Darwinian struggle for existence.
But as is the wont of human brain, brighter of the lot,  dealing with this  incipient subject of numbers and shapes,  perceived interesting patterns in some of these entities,  their  interrelations, and posed interesting questions that entailed  concepts  such as the  zero, decimal system  (an Indian gift),  prime numbers, irrational numbers,  the Pythagorean relation between sides and the hypotenuse of  right triangles (described also in the {\it Baudhayana sutras} \cite{m4}), and so on.
 
Playing and tinkering  are  natural human instincts  and, not surprisingly therefore,  curious and   innovative minds  toyed around with  patterns found among the   arithmetical  and geometrical  entities to create  rich logical  systems from which one  could establish (i.e. prove) non-trivial results (i.e. theorems) such as  the number of prime numbers being infinite or the Pythagoras theorem, by deploying  imaginative and clever tricks  on a set of very few, almost self-evident  assumptions (i.e. axioms) and by making use of logical rules of operations. 

The Midas touch of  mathematical minds  gave   fillip  to  axiomatic systems, like for example Euclid's geometry,  to grow wings as though of their own and to fly out to magical and intangible worlds. Numbers and geometrical concepts  got  transmuted and generalized to ever increasingly abstract but beautiful creatures, seemingly far removed from concrete reality. New kinds of numbers were imagined. Genie of mathematics was out of the bottle (Klein's?)!

\section*{Abstraction and Elegance}

Creativity in mathematics displays not only use of abstract concepts and procedures but also elegance and beauty. Puzzling and non-trivial conundrums are often spotted or conjured up, and   are  eventually settled  in  ingenious ways. As an illustration, consider the hypothesis H that  any irrational number raised to the power of an irrational number is always irrational. Is H  true?

Some mathematicians used a brilliant argument to prove that the above hypothesis is false. Take the case when $x = \sqrt {2}$. So, x  is manifestly irrational. Construct a new  number  y,
  $$y \equiv x^x \ .$$  
  Now, clearly if $y$ is rational, the hypothesis H is incorrect, and we are done. On the other hand, if $y$ is irrational (as  H will have us believed),  define another   number,
 $$z \equiv  y^x \ .$$
Then,
$$z=  (x^x)^x = x^{x^2}=(\sqrt{2})^2=2 \ ,$$
which is rational. 

Thus, using the above counter example(s) one has  disproved the hypothesis. But curiously, the preceding argument is completely silent as to  whether $y$ is rational or not. All it demonstrates is that if $y$ is irrational then $z=y^x$ is not, when $x = \sqrt {2}$.

Speaking of abstraction, the imaginary number $i \equiv \sqrt{-1}$ (iota)  arose  as a mathematical device, representing an abstract solution of the quadratic equation $x^2 + 1 = 0$. It is interesting to note that we can give a  geometrical meaning to iota (as the altitude)  by interpreting $x^2 + 1 = 0$ to be a  Pythagorean  relation for an abstract right triangle having a unit base length but  a hypotenuse of zero length!

  Complex numbers, constructed out of a combination of  real numbers and iota, gave birth to a rich structure consisting of  elegant theorems involving analytic functions, conformal mappings, analytic continuation, Fourier transforms, Riemann zeta functions, etc.   Before the arrival of iota, it was  unimaginable that geometrical   entities and  functions related to logarithm could get linked up, as in the case of the amazing formula due to Euler,
 $$ e^{i \theta}=\cos {\theta} + i \sin {\theta}\ . $$
 Complex number was only the beginning. Soon, it was raining quaternions and  Grassmann variables!

Till the advent of quantum theory, complex numbers were thought of only  as useful tools, having no correspondence    with the actual world. After all, in Nature, measurable quantities  are always real. But   post 1920s, it was realized that the physical world is quantum mechanical in nature, in which the Schr\"{o}dinger equation,
$$ i \hbar \frac {\partial \psi} {\partial t} = \hat {H} \psi \eqno(1)$$
 determines the time evolution of a physical system, given the Hamiltonian $\hat{H}$, while  the commutator bracket of position and momentum operators,
$$[\hat{x}, \hat{p}] = i \hbar \eqno(2)$$
is responsible for Heisenberg's uncertainty principle $\Delta x \Delta p \geq \hbar /2$.
Eqs.(1) and (2) emphasize  that $i \equiv \sqrt{-1}$ indeed `exists' in the real world, and it is probably as real as time is \cite{m15}. Quantities measured in experiments are still real, as they invariably  correspond to eigenvalues of  hermitian operators representing the physical observables.

To cite another case, the measure of  distance between any two points  that relied on Pythagoras theorem  in the standard  Euclidean geometry  got generalized to abstract ones involving metric tensors suitable for curved or warped spaces described by non-Euclidean geometry. Pioneers of these developments  were titans  like Gauss, Lobachevsky, Bolyai, Riemann, David Hilbert, Poincare and others. For a   comprehensive account of the historical development of the field as well as the   monotonic rise of importance of abstract mathematics  in physics, readers are referred to N. Mukunda's recent article\cite{m6}.

 A tiny portion of the subject of non-Euclidean geometry came handy when Einstein  formulated a relativistic theory of gravity (i.e. general relativity) during 1907-1915.   General relativity took geometry of the four dimensional space and time to an exalted level, where it became as much dynamical as  the  matter itself, whose energy and momentum caused the geometry to be non-Euclidean \cite{m9, m16}.  
 
 An oft raised question is  whether  human beings were genetically  programmed to be abstract mathematicians/theoretical physicists.  Instead,  suppose we ask:  were we genetically wired to have been swayed emotionally by sophisticated music? Putting  forward a thesis  that musical melodies have their  roots in the sequence of notes present in bird songs, and that those early humans who were sensitive to and were drawn to simple jingle of a koel's cooing or of other singing birds,  had greater chances of survival, finding their mates  as well as  passing on their genes to offspring (since  birds flocked in regions where water and food are  abundant), one can  possibly explain why music affects us emotionally, and why  its  primitives  are  similar to bird songs.

 And then with time,  simple tunes grew in richness because of the fascinating flexibility of  brain which grows  more  neuronal connections  with extra stimuli provided by  inputs and outputs from other musically minded people,  leading to further creative churnings in music. The ever increasing complex networks  both of inter-neuronal highways in the brain as well as of musicians entailed augmentation of a  sophisticated body of music. Contemporary  music  is obviously far more complex and richer  in comparison to the brief melody of  a bird song.
 It will not be far fetched to expound a similar theory  in the case of mathematics for  its ever growing complexity and sophistication. So, what started with an evolutionary advantage, became richer, more abstract  and multi-layered over time.

\section*{Hilbert, G\"{o}del and `Hotel Infinity'}   

David Hilbert, an all time giant of a mathematician, envisaged an axiomatic formulation of  mathematics, inspired significantly by the mathematician Moritz Pasch's work, in which all elements of mathematics  are divorced entirely  from pictures, concrete objects, physical reality, etc., in order that mathematics becomes  purely  a formal and abstract system so that, starting from a set of axioms, one could prove or disprove all well formed mathematical statements \cite{m7}.

It is interesting to note that in the celebrated list of 23 `superproblems' which Hilbert had presented in a lecture at Paris in 1900, the 6-th problem was about axiomatic treatment of physics, since its  fundamental laws  are expressed in the language of mathematics \cite{m11}.
But his program of a complete axiomatic formulation of mathematics received a jolt when G\"{o}del came up with his incompleteness theorem which, simply stated,  implies that  in a consistent logical system  there will exist  well formed logical expressions that  can neither be proved nor disproved \cite{m7,m8}.  
 
At this point, one  wonders  about the effect of G\"{o}del's incompleteness theorem on physical theories which lean heavily on mathematics.
Does ambiguity in the physical world enter through a G\"{o}delian back door? Or, is it that whenever an undecidable statement springs up in a physical theory via mathematics, one simply subjects it to an experimental test in order to obtain its truth value?   For, natural science has the luxury of experimentation! 

 Hilbert in 1925 had conjured up a hotel consisting of countable infinity of rooms, bearing room numbers n = 1, 2,..., $\infty $, in order to illustrate the  counter-intuitive nature of a set containing infinite number of elements. He had argued that even if this hotel is fully occupied initially, he could still accommodate  M  unexpected visitors. All he would  do is to   shift the existing occupant of  room number  n to M+n. Then, the new guests could move into the first M vacant  rooms. Hilbert went on to claim that even when a countable infinity of  visitors arrived suddenly, he could repeat the feat.  He would now move the old occupant of room n to 2n, so that an infinity of rooms, bearing odd  numbers as room numbers, would lie vacant so that  just  arrived  guests (infinity of them) could  move in. 
 
 One could possibly extend  Hilbert's argument also for an uncountable infinite set  (e.g. a set that has the same cardinality as the set of  real numbers)   trivially  using a concrete example. Consider a semi-infinite rod  inside a semi-infinite tube, with their ends matching, so that there is no empty space inside the tube. One can create as much space in the tube as one wishes simple by pushing the end of the rod so that it slides forward, while keeping the end of the  tube fixed.   
Apparently George Gamow played a pivotal role in elevating Hilbert's `hotel infinity' to five star status as the former  used it to explain the incessant expansion of an infinite space in the context of big bang model of expanding universe\cite{m10}.

 One may speculate about another possible consequence of `hotel infinity'. Dirac's theory of electrons and positrons envisages that the vacuum is a `sea of electrons' filling up all the negative energy states. Suppose there is a physical mechanism (like  applying a strong electric field) to shift  every electron from its negative energy level to the next lower negative energy state simultaneously so that Pauli's exclusion principle is not violated.   Then,  Hilbert's  argument will create yet another difficulty for the Dirac sea, since moving negative energy electrons to lower levels will entail +ve energy electrons to fall on to the vacant -ve energy levels releasing gamma photons. It will trigger an unending avalanche of electrons making transitions to more -ve energy states, all heading towards the bottomless pit and generating high energy photons continuously. This  makes the  physical vacuum unstable, which is  an undesirable feature.   

Incidentally, it was Hilbert, who proposed an action for  general relativistic gravity from which one could  also derive, using the principle of least action,  Einstein's equations that relate geometry to matter \cite{m9}. Another serendipitous  Hilbert-physics  connection  is that according to quantum theory, for every physical system there corresponds   an infinite dimensional  linear vector space, with complex numbers as the field, which is endowed with an inner product -  in short,  an example of a Hilbert space! 

\section*{Physical Sciences and the Unreasonable Effectiveness of Mathematics}
 
 Modern mathematics  is not  a  recondite recreation akin to the game of chess (yet another Indian innovation). It is an established fact that  all fundamental laws of Nature happen to be  expressed in mathematical language. As early as 1959, Eugene Wigner, an eminent theoretical physicist, drove home the point about   
 `The unreasonable effectiveness of mathematics in the Natural sciences' \cite{m12} .
 Abstract concepts and relations like  tensors, affine connection,  Hilbert space, operators,  matrices, commutators, Lie algebras and groups, Grassmann variables and many more, created by  mathematicians for  their  own sake, turn out to describe fundamental truths concerning  the actual universe.
 
  General relativity and 
 the bizarre quantum world of electrons, photons, atoms, etc. bear testimony to the fact that most of the aforementioned abstract concepts  `exist' in the real world. But that is perplexing, as pointed out by Wigner so eloquently, for they come out of mathematicians' brains! However,  a crucial fact has been overlooked -   only a miniscule portion of the vast body of mathematical creations  gets to enjoy the status of being the language of fundamental laws of Nature. The supply of abstract concepts from mathematicians are incredibly larger than the demand from physics. Therefore, should one  be surprised that a small subset of mathematical  ideas based on logic, beauty, elegance and abstract generalization of concepts, which were rooted to reality once upon a time, gets realized  in fundamental physics?

Any idea whose effect or consequences  cannot be measured, in principle, is not part of science. It so happens that  physical entities  are measured quantitatively (i.e. in terms of numbers),  and hence, it is natural that  their interrelations,  including temporal cause and effect connections,  better be based on a language that is numeric, precise,  logical and  unambiguous.   Mathematics  is  such a  language and thus,  is ideally  suited for expressing laws and principles in  physical sciences. 

Furthermore, since mathematics deals with  entities, relations and theorems that are abstract in character, it is naturally  amenable to wider applications. For, one  can substitute  a physical concept in place of an  abstract entity, use the established  mathematical machinery and  arrive at a non-trivial result that is concrete, provided the logical structure of the abstract system  is `isomorphic' to the basic framework of the concrete system. In other words, mathematical modeling of a physical system is often a case of moving from general to particular.  To pin point the counter-intuitive nature of any infinite mathematical set that is precisely what Hilbert had done with his example of hotel rooms and guests or what Gamow had attempted when he considered the infinite physical space as an instance of an infinite set. By its very construction, mathematics encompasses such diverse, generalized   as well as  abstract elements and structures that it has a greater propensity to act as language for physical sciences.

 The real world has  continued to inspire  mathematicians, be it the birth of  calculus for finding trajectories of bodies moving  in gravitational fields or of distribution theory that ensued from  Dirac delta function. It is common wisdom that  systematic analysis of gambling outcomes  by Fermat, Pascal and Huygens had ushered in the mathematical theory of probability.  In a lighter vein, the Mahabharata hero  Yudhisthira could have  outwitted Shakuni  in the game of dice had he brushed up on the theory of chances. 

One could speculate whether the chance coincidence of angular sizes of  the Sun and  Moon being almost same at   present (and also during the historical period)  causing  eclipses to occur, played a significant role in the development of mathematics in the past.  After all,  mathematicians of great repute  exercised their minds to understand and predict  eclipses, be it Hipparchus, Aryabhatta or Varahamihira. In other words,   it is scientifically relevant  to ask  whether  in the absence of Moon eclipsing the Sun, there would  have been enough impetus and motivation to develop trigonometry and other useful  computational techniques  \cite{m14}.  

The renowned astrophysicist S. Chandrasekhar arrived at many fundamental truths about gravitating systems in astronomy, employing beautiful and elegant mathematical analysis. His predictions concerning upper limit on white dwarf  mass, dynamical friction, magnetohydrodynamics instabilities, etc. stood the tests of observational verification, vindicating the power of mathematical rigor\cite{m13, m17}. 

Not needing the crutches of  experiments, mathematicians rely purely on logical deductions and their thinking prowess,  celebrating thereby {\it Cogito Ergo Sum} of Descartes.  Descartes had amalgamated algebra and geometry to create coordinate geometry. So, on a lighter note, we may juxtapose the utterances of Descartes and Archimedes, to envisage a   maxim {\it `Cogito Ergo Eureka’} for the pursuit of mathematics! 

\begin{acknowledgments}
It is a pleasure to thank R. Ramaswamy, Srinivas Rau, B. Sury  and Jugal Kishore Verma for their useful comments.
\end{acknowledgments}
%
%

\end{document}